Brief Communication

**From P100 to P100': Conception and improvement of a new citation-rank approach in bibliometrics**

Lutz Bornmann[1], Rüdiger Mutz[2],

[1] Division for Science and Innovation Studies, Administrative Headquarters of the Max Planck Society, Munich, Germany, Email: bornmann@gv.mpg.de, Tel: +49 89 2108 1265

[2] Professorship for Social Psychology and Research on Higher Education, ETH Zurich, Zurich, Switzerland, Email: ruediger.mutz@gess.ethz.ch, Tel: +41 44 632 4918

Abstract:

Properties of a percentile-based rating scale needed in bibliometrics are formulated. Based on these properties, P100 was recently introduced as a new citation-rank approach (Bornmann, Leydesdorff, & Wang, in press). In this paper, we conceptualize P100 and propose an improvement which we call P100'. Advantages and disadvantages of citation-rank indicators are noted.

Keywords:

Percentile; P100; P100'; Citation-rank

**The use of reference sets and properties of a percentile-based rating scale needed in bibliometrics**

Cross-field and cross-time-period comparisons of citation impact for research evaluation purposes are only possible if the impact is normalized (standardized) (Bornmann & Marx, 2013; Schubert & Braun, 1986). For its citation impact to be normalized, a paper needs to have a reference set: all the papers published in the same publication year and subject category. Percentiles have been proposed as an alternative to normalization on the basis of central tendency statistics (arithmetic averages of citation counts) (Bornmann, Leydesdorff, & Mutz, 2013; Bornmann & Mutz, 2011; Bornmann, Mutz, Marx, Schier, & Daniel, 2011; Schreiber, in press). Percentiles are based on an ordered set of publications in a reference set, whereby the fraction of papers at or below the citation counts of a paper in question is used as a standardized value for the relative citation impact of this focal paper. This value can be used for cross-field and cross-time-period comparisons. If the normalized citation impact for more than one paper is needed in a research evaluation study, this percentile calculation is repeated (by using corresponding reference sets for each one).

In our opinion, a percentile-based rating scale allowing cross-field and cross-time-period comparisons should have the following properties:

- The paper with the highest citation count in a reference set should receive the highest scale value (percentile). If there is more than one most highly cited paper, all these papers should be attributed to the same highest scale value.
- The paper with the lowest citation count or zero citations in a reference set should receive the lowest scale value. If there is more than a single most lowly cited paper, all these papers should be attributed to the same lowest scale value.
- Papers with the same citation impact in a reference set should receive one and the same scale value. The problem of ties should be solved unambiguously.
- In a reference set, the scale values should be distributed from 0 to 100 exactly and should be comparable across different reference sets. The paper with the highest impact (lowest impact) in one reference set should receive the same scale value as the paper with the highest impact (lowest impact) in another reference set. Since all most highly cited (all least cited) papers in their corresponding reference sets have the same top (lowest) performance, they should receive the same scale value.
  In the percentile-based approaches proposed hitherto (see overviews in Bornmann, et al., 2013; Waltman & Schreiber, 2013), papers with the highest impact (lowest impact) receive different percentiles depending on the properties of the corresponding reference sets.
- Based on the scale values, it should be possible to categorize the papers in question into a small set of (percentile) rank classes.
- Aggregation of scale values should be possible, e.g., across a publication set for a scientist or an institution.
- The Centre for Science and Technology Studies in Leiden (CWTS) uses weights in its percentile-based approach to categorize publications to $PP_{top\ 10\%}$ (Waltman et al., 2012). $P_{top\ 10\%}$ is the number and $PP_{top\ 10\%}$ the percentage of papers which belong to the 10% most highly cited papers in the corresponding reference sets. Weights for papers should be avoided in a percentile (scale value) calculation, since the statistical analysis of publication sets with weights is far more complicated than without weights.

**Conceptualization of P100 and its refinement to P100'**

Recently, P100 was introduced as a new citation-rank approach to conform as precisely as possible to the above properties (Bornmann, et al., in press). In this paper, P100 is conceptualized and refined further on to P100'.

Citations of papers in a reference set can be ranked according to their frequencies of papers, which result in a so-called size-frequency distribution (Egghe, 2005). This distribution can be used to generate a citation-rank where the frequency information is ignored. In other words instances of papers with the same citation counts are not considered. This perspective on citation impact focuses on the distribution of the unique citation counts with the information of maximum, medium, and minimum impact and not on the distribution of the papers (having the same or different citation impact) which is the focus of interest in the conventional citation analysis (Bornmann, et al., in press).

Insert Figure 1 here

Figure 1 shows a comparison of distributions based on citations for all papers (left side) and on unique citations (right side) in chemistry, physics, and psychology. For each field, all the articles from 1990 published in the corresponding subject categories (e.g., "physics, applied," "physics, nuclear," etc.) were downloaded in June 2013 from Web of Science (Thomson Reuters). The citations have been logarithmized (log(cit+1)) for presentation in the figure. It is clearly visible for all the fields in Figure 1 that the use of unique citations leads to a great reduction of the visualized data. For example, there are 54,564 articles in the chemistry-related subject categories. The use of unique citations leads to only 379 values. Whereas the distribution of the (logarithmized) citation counts for all papers is skewed to the right, the unique citations are distributed symmetrically, nearly log-normal.

To generate citation-ranks for a reference set, the unique citations are ranked in ascending order from low to high citation counts and ranks are attributed to each citation count with rank 0 for the paper with the lowest impact or zero citations. In order to generate values on a 100-point scale (P100), each rank i is divided by the highest rank $i_{max}$ and multiplied by 100, i.e.

$100*(i/i_{max})$

Insert Table 1 here

This procedure is an ordinal or monotone transformation of the original citation data. The citation rank reflects the relative rank of a paper within its citation-impact reference set. The citation-rank approach is illustrated in a numerical example (Table 1). In contrast to the traditional size-frequency distribution, the citations are not sorted by the frequency but by the size of citations in ascending order. For a paper with 4 citations, a P100 value of 66.67 is calculated, for example, whereas the percentile value is 77.78.

Most statistical software (e.g., SAS, STATA) offers ranking procedures which can be used for the calculation of citation-rank values (SAS Institute Inc., 2008; StataCorp., 2011).

Insert Table 2 here

As two examples of similar reference sets in Table 2 show, P100 has undesirable properties which should be avoided. Each reference set consists of six publications with the following numbers of citations: (1, 2, 3, 4, 5, 6) and (1, 2, 2, 4, 5, 6). In the first reference set, the publication with 4 citations has a higher P100 value (P100=60) than in the second reference set (P100=50). This cannot be justified, because 3 out of 6 papers have fewer than 4 citations in both reference sets.

To avoid this paradoxical situation in which the scale value of a paper can increase as a result of the fact that another paper receives an additional citation, we developed P100' (see Table 2). In contrast to P100, the ranks for P100' are not only based on the unique citation distribution, but consider also the frequency of papers with the same citation counts. For P100', each rank j is divided by the highest rank $j_{max}$ or (n-1) papers in the reference set and is multiplied by 100, i.e.

$$100*(j/j_{max})$$

N papers with the same citation receive the same lowest rank value $r_0$, whereas the next paper with a higher citation value receives a rank value $r_1$ of $r_0+n$. For example, for a publication set with citation counts of 10, 12, 30, 30, 40, 50 the rank values are 0/5, 1/5, 2/5, 2/5, 4/5, 5/5.

Whereas in Table 2 (second reference set) rank i of the paper with 4 citations is 2 (which is based on the unique citation distribution), it increases to rank j=3 if papers with the same citation counts are further taken into account. As a result of this change of the ranking procedure, both papers with 4 citations receive the same scale value (P100'=60).

Insert Table 3 here

Table 3 shows P100' values for the same reference set as presented in Table 1. For the paper with 4 citations, a P100' value of 50 is calculated, for example, whereas the percentile value is 77.78. The P100' value is not only different from the percentile, but has also another meaning. The P100' value of 50 means that 50% of all citation-rank places are better than this rank place. In contrast, the percentile value of 77.78 means that 7 papers of all 9 papers (=77.78%) in the reference set have 4 citations or less than 4 citations.

**Advantages and disadvantages of the citation-rank approach**

The advantages of the citation-rank approach:

- *Fixed scale:* The paper with the highest citation impact always receives the same value of 100 (independently of the distribution of citation counts in different reference sets). The same is true for the paper with the lowest citation impact (mostly the value of zero citations).
- *Simple treatment of ties*: The use of the size-frequency distribution instead of all citation counts avoids complicated methods for dealing with ties (e.g., the fractional counting approach of Waltman & Schreiber, 2013).
- *Expected value*: The expected value or mean value of P100' values in a reference set amounts to 50%.
- *Rank classes*: Based on the scale values, it is possible to categorize the papers in question into a set of rank classes.
- *Ranking information*: Ordinal information as a ranking (e.g., x is better than y) is more satisfactory in (scientific) achievement assessments than continuous grades (e.g., x is 5 points better than y) while making fewer demands on data quality. The most quantitative information in daily life, as grades or ratings, is ordinal by nature. In examination systems, for instance in England and Wales, the reporting is by grades from A* to E, which are definitely ranks (Fielding, Yang, & Goldstein, 2003).
  The "ranking information" advantage is shared by citation-rank and percentile-based approaches.

- *Easy calculation*: The calculation of citation ranks is very easy. All the papers with the same citation counts are eliminated with a sorting procedure. A ranking procedure can be applied to this transformed citation data in order to calculate the citation rank (e.g., the percent option in SAS).

There are also some limitations to the citation rank approach:

- *Interpretation problems*: Compared to citation-rank percents, percentiles allow direct interpretation of the data in the sense that the $x^{th}$ percentile of a paper means that *x* percent of all papers in the reference set fall below the respective paper. This interpretation is almost impossible for citation ranks.
- *Lost scale property*: The scale values generated by the citation-rank approach have lost their interval-scale property. Any monotone transformation (e.g., $x^2$) is possible and does not change the scale.
- *Same number of citations*: If all papers in a reference set have the same number of citations, P100 and P100' cannot be calculated. The unique citation distribution consists of only one value and there are not any citation impact differences between the papers.
- *Reliability problems*: Many papers with the same citation counts in a reference set make the citation value and the corresponding rank value more reliable, in the sense of stability against random fluctuations, than citation counts based on only one or two papers. Many papers with the same number of citations can reflect the citation-impact level in a reference set better than a few papers with infrequent citation counts (e.g., the most frequently cited papers). Further methodological research is necessary to include reliability and standard error issues in the citation-rank approach.

**Acknowledgements**

We are grateful to Michael Schreiber for exchanges of ideas and to an anonymous reviewer for recommendations to significantly improve the paper.

**Figure**

Figure 1. Comparison of distributions based on citation counts for all papers (logarithmized, left side) and unique citation counts (logarithmized, right side) in chemistry, physics, and psychology

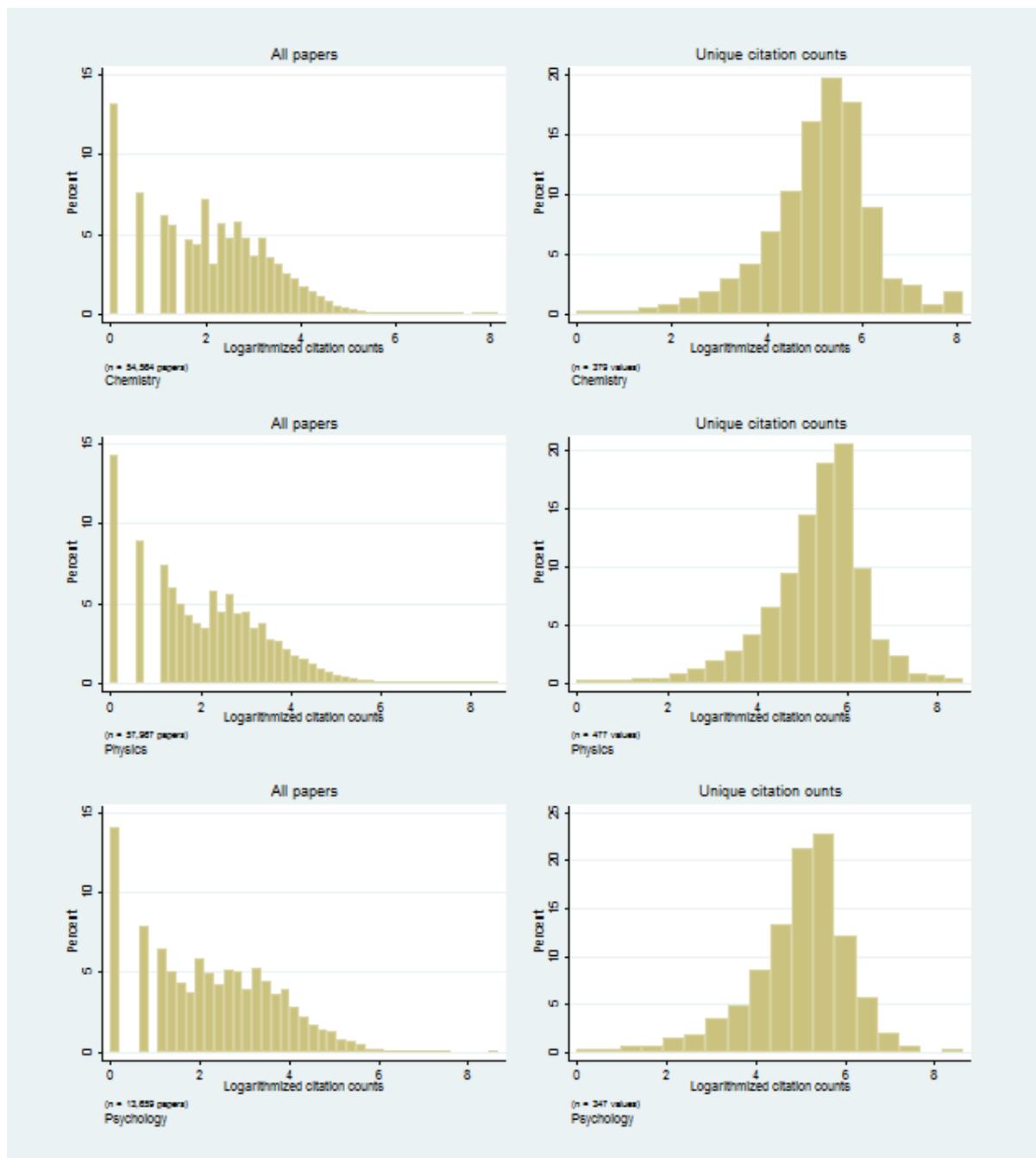

**Tables**

Table 1. The use of the size-frequency distribution of citations to calculate P100 in comparison to traditional percentiles or cumulative frequencies in percent (numerical example, N=9 papers in a reference set)

| Citations | Number of papers | Rank i (citations) | P100: citation-rank $100*(i/i_{max})$ | Percentile (cumulative frequencies in percent) |
|---|---|---|---|---|
| 0 | 1 | 0 | 0 | 11.11 |
| 1 | 1 | 1 | 16.67 | 22.22 |
| 2 | 1 | 2 | 33.33 | 33.33 |
| 3 | 1 | 3 | 50.00 | 44.44 |
| 4 | 3 | 4 | 66.67 | 77.78 |
| 7 | 1 | 5 | 83.33 | 88.88 |
| 10 | 1 | 6 | 100.00 | 100 |

Table 2. Two examples of similar reference sets to demonstrate the problem of P100 (N=6 papers in a reference set). In the first reference set, P100 and P100' values are identical.

First reference set

| Citations | Number of papers | Rank i (citations) | P100 and P100': citation-rank $100*(i/i_{max})$ |
|---|---|---|---|
| 1 | 1 | 0 | 0 |
| 2 | 1 | 1 | 20 |
| 3 | 1 | 2 | 40 |
| 4 | 1 | 3 | 60 |
| 5 | 1 | 4 | 80 |
| 6 | 1 | 5 | 100 |

Second reference set

| Citations | Number of papers | Rank i (citations) | P100: citation-rank $100*(i/i_{max})$ | Rank j (citations and papers) | P100': citation-rank $100*(j/j_{max})$ |
|---|---|---|---|---|---|
| 1 | 1 | 0 | 0 | 0 | 0 |
| 2 | 2 | 1 | 25 | 1 | 20 |
| 4 | 1 | 2 | 50 | 3 | 60 |
| 5 | 1 | 3 | 75 | 4 | 80 |
| 6 | 1 | 4 | 100 | 5 | 100 |

Table 3. The use of the size-frequency distribution of citations to calculate P100' in comparison to traditional percentiles or cumulative frequencies in percent (numerical example, N=9 papers in a reference set)

| Citations | Number of papers | Rank j (citations and papers) | P100': citation-rank $100*(j/j_{max})$ | Percentile (cumulative frequencies in percent) |
|---|---|---|---|---|
| 0 | 1 | 0 | 0 | 11.11 |
| 1 | 1 | 1 | 12.5 | 22.22 |
| 2 | 1 | 2 | 25.0 | 33.33 |
| 3 | 1 | 3 | 37.5 | 44.44 |
| 4 | 3 | 4 | 50.0 | 77.78 |
| 7 | 1 | 7 | 87.5 | 88.88 |
| 10 | 1 | 8 | 100.00 | 100 |